%% file: acl_latex.tex
\PassOptionsToPackage{most}{tcolorbox} 
\documentclass[11pt]{article}

\usepackage[final]{acl}

\usepackage{times}
\usepackage{latexsym}

\usepackage[utf8]{inputenc}
\usepackage{amsmath,amssymb}
\usepackage{pifont}
\usepackage{placeins} 
\usepackage{booktabs}
\usepackage{multirow}
\usepackage{lipsum}
\usepackage{tabularx}
\usepackage{subcaption}
\usepackage{arydshln}
\usepackage{adjustbox} 
\usepackage{makecell}

\usepackage{siunitx}
\usepackage[T1]{fontenc}

\usepackage{inconsolata}
\usepackage{dblfloatfix}
\usepackage{graphicx}
\usepackage{enumitem}
\usepackage{cuted}
\usepackage{capt-of} 
\usepackage{graphicx}
\usepackage{dblfloatfix} 
\usepackage{caption}  
\usepackage{fontawesome5} 
\usepackage{hyperref}     
\usepackage{xcolor}       

\title{MemGovern: Enhancing Code Agents through Learning from\\ Governed Human Experiences}

\author{
\textbf{Qihao Wang\textsuperscript{1*}},
\textbf{Ziming Cheng\textsuperscript{2*}},
\textbf{Shuo Zhang\textsuperscript{10*}},
\textbf{Fan Liu\textsuperscript{3*}},
\\
\textbf{Rui Xu\textsuperscript{4}},
\textbf{Heng Lian\textsuperscript{5,10}},
\textbf{Kunyi Wang\textsuperscript{6,10}},
\textbf{Xiaoming Yu\textsuperscript{7}},
\textbf{Jianghao Yin\textsuperscript{3}},
\textbf{Sen Hu\textsuperscript{8,10}},
\textbf{Yue Hu\textsuperscript{1}},
\\
\textbf{Shaolei Zhang\textsuperscript{9\dag}},
\textbf{Yanbing Liu\textsuperscript{1\dag}},
\textbf{Ronghao Chen\textsuperscript{8,10\dag}},
\textbf{Huacan Wang\textsuperscript{1,10\dag}}
\\
\\
 \textsuperscript{1}UCAS,
 \textsuperscript{2}NUS,
 \textsuperscript{3}ECNU,
 \textsuperscript{4}FDU,
 \textsuperscript{5}XDU,
 \textsuperscript{6}UBC,
 \textsuperscript{7}HKUST(GZ),
 \textsuperscript{8}PKU,
 \textsuperscript{9}RUC, 
 \textsuperscript{10}QuantaAlpha
\\
\small {
    \textbf{\textsuperscript{*}These authors contributed equally to this work.}
}
\\
 \small{
   \textbf{\dag Correspondence:} 
   \href{mailto:zhangshaolei98@ruc.edu.cn}{zhangshaolei98@ruc.edu.cn},
   \href{mailto:liuyanbing@iie.ac.cn}{liuyanbing@iie.ac.cn},
   \href{mailto:chenronghao@alumni.pku.edu.cn}{chenronghao@alumni.pku.edu.cn},
   \href{mailto:wanghuacan17@mails.ucas.ac.cn}{wanghuacan17@mails.ucas.ac.cn}
 }
}

\begin{document}
\maketitle

\begin{center}
    \vspace{-15pt}
    \large
    \faGithub\hspace{6pt}\href{https://github.com/QuantaAlpha/MemGovern}{\texttt{\color{black}https://github.com/QuantaAlpha/MemGovern}}
\end{center}

\vspace{10pt} 

\begin{abstract}
While autonomous software engineering (SWE) agents are reshaping programming paradigms, they currently suffer from a “closed-world” limitation: they attempt to fix bugs from scratch or solely using local context, ignoring the immense historical human experience available on platforms like GitHub. Accessing this open-world experience is hindered by the unstructured and fragmented nature of real-world issue-tracking data. In this paper, we introduce MemGovern, a framework designed to govern and transform raw GitHub data into actionable experiential memory for agents. MemGovern employs experience governance to convert human experience into agent-friendly experience cards and introduces an agentic experience search strategy that enables logic-driven retrieval of human expertise. By producing 135K governed experience cards, MemGovern achieves a significant performance boost, improving resolution rates on the SWE-bench Verified by 4.65\%. As a plug-in approach, MemGovern provides a solution for agent-friendly memory infrastructure.

\end{abstract}

\begin{figure}[h]
    \centering
    \includegraphics[width=0.95\textwidth]{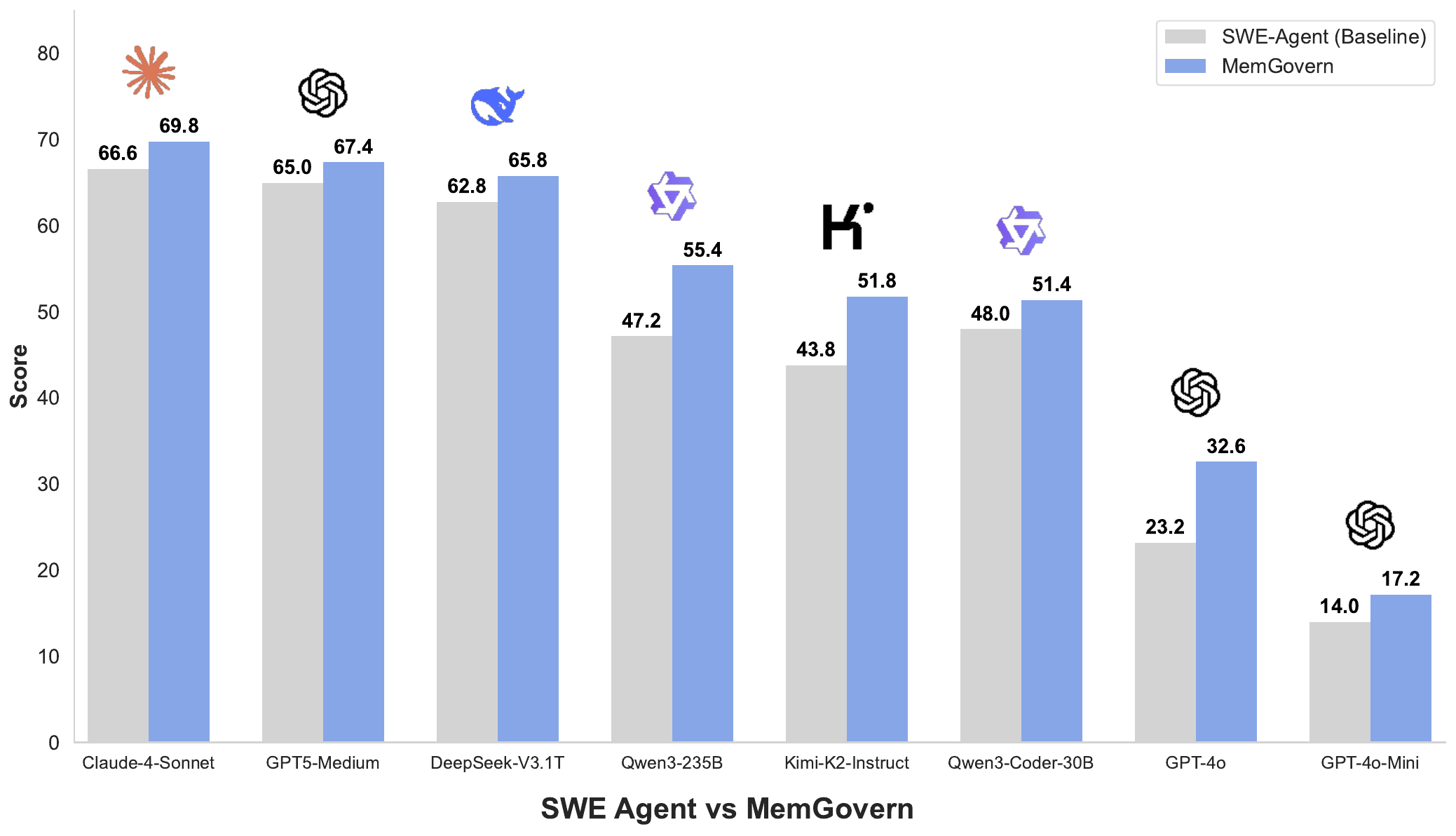}\hspace*{20pt}
    \vspace{-5pt}
    \caption{Performance comparison of SWE-Agent(the base framework) and MemGovern across different LLM backbones on SWE bench Verified.}
    \label{fig:teaser} 
\end{figure}

\newpage

\section{Introduction}

With the remarkable progress of large language models (LLMs) in code understanding and generation \citep{chen2021codex, li2023starcoder, roziere2023codellama, achiam2023gpt4}, autonomous software engineering agents (a.k.a, Code Agents) are reshaping the paradigm of programming \citep{hong2023metagpt, qian2023chatdev, wu2023autogen}. Among such tasks, autonomously fixing bugs in GitHub repositories has emerged as a key benchmark for evaluating code agent capabilities and became a central research focus \citep{jimenez2024swebench, yang2024sweagent, liu2024repobench}.

In real-world software engineering practice, developers rarely fix bugs from scratch \citep{ko2006exploratory}. When confronted with complex issues, engineers typically search collaborative platforms such as GitHub to examine how similar problems were addressed in the past \citep{sadowski2015developers, xia2017measuring}. These records on GitHub inherently encode expert debugging reasoning and repair patterns \citep{rahman2014insight, tian2018learning}. Ideally, an advanced code agent should be able to exploit such open-world experience like human developers, leveraging historical repair strategies to improve its reasoning depth and accuracy when handling difficult bugs \citep{shinn2023reflexion}.

\begin{figure*}[h]
  \centering
  \includegraphics[width=0.8\textwidth]{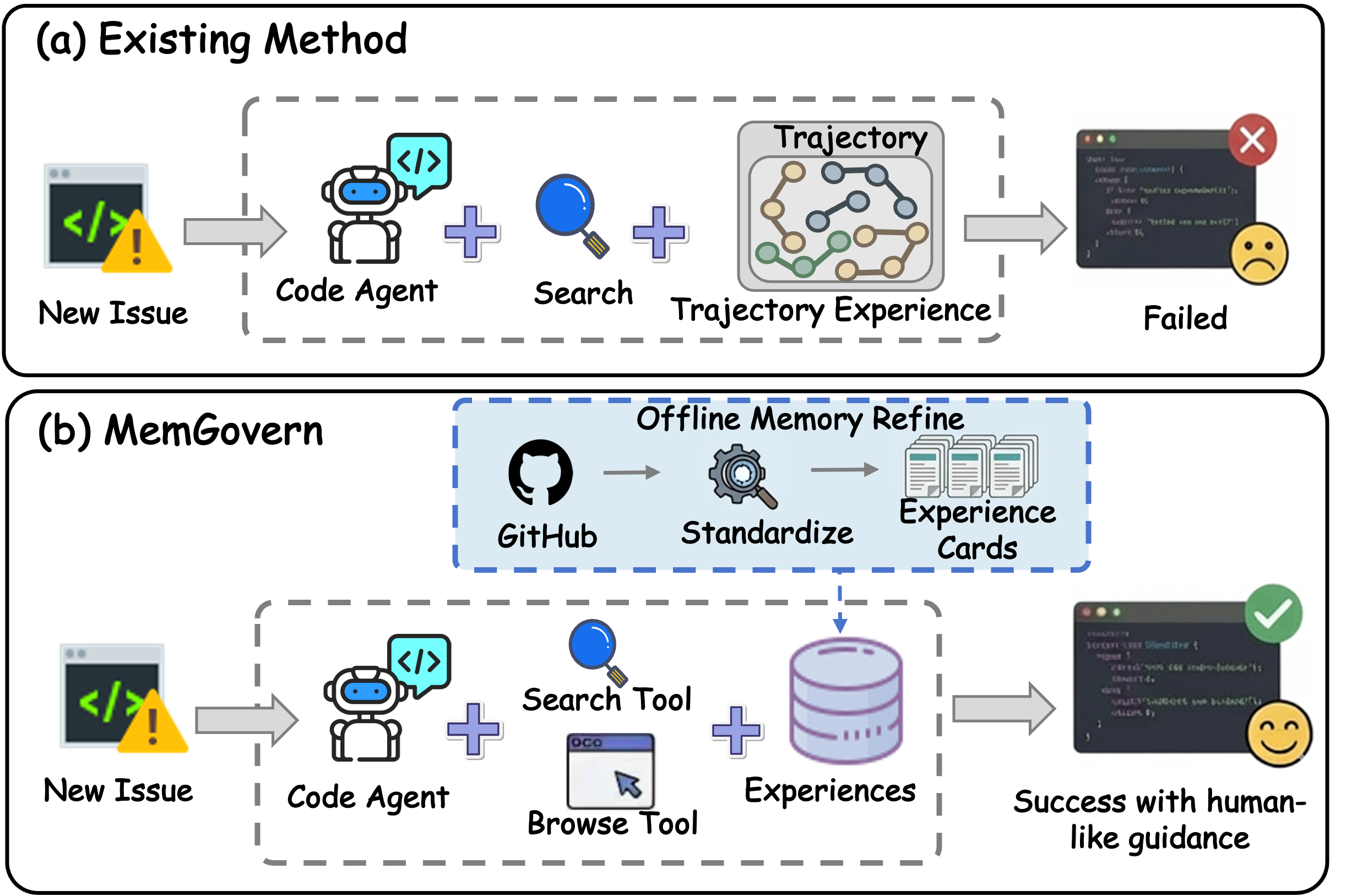}
  \caption{Comparison of MemGovern with existing methods. MemGovern learns from human experience by governing raw data into agent-friendly memories.}
  \label{fig:vdr_pipeline}
\end{figure*}

\textbf{\emph{Although GitHub holds a vast repository of human experience, converting it into agent-friendly knowledge remains challenging.}} First, raw issue and pull request discussions contain large amounts of many unstructured and fragmented information, such as social exchanges and procedural communication, which often obscures technical insights \citep{kalliamvakou2014promises, bird2009fair}. Second, variations in terminology, module organization, and coding styles across projects hinder the standardized transfer of repair knowledge \citep{allamanis2018survey}. Overall, cross-repository human issue experience on GitHub is both noise-dense and highly heterogeneous, lacking an effective governance mechanism to convert it into retrievable and verifiable knowledge representations. This limitation constitutes a primary bottleneck that confines most approaches to within-repository retrieval \citep{jimenez2024swebench}.

To address these challenges, we propose MemGovern, a governance framework that transforms human experience into agent-friendly experiential memory, thereby providing code agents with an \textbf{\emph{experience infrastructure}}. MemGovern aims to transform disorganized GitHub repair records into structured memories that can be efficiently exploited by agents \citep{park2023generative, pack2024memgpt}. MemGovern introduces \textbf{\emph{experience governance}} to automatically clean and standardize cross-repository experience, organizing it into standardized experience cards that capture key dimensions such as modification strategies and validation methods. Accordingly, MemGovern introduces \textbf{\emph{agentic experience search}}, which enables agents to interact with the experiential memory through multiple rounds of searching and browsing, closely mirroring how human engineers explore prior cases \citep{singer2010software}. This design allows agents to identify the underlying repair logic in similar examples rather than relying solely on shallow semantic matching. Specifically, MemGovern is designed as a plug-and-play module that can be seamlessly integrated into existing agent scaffolds with minimal modifications. In our implementation, we adopt the state-of-the-art SWE-Agent as the shared backbone. Our experimental results on the SWE-bench demonstrate that MemGovern improves the ability of code agents to resolve real-world bugs by 4.65\% on average, owing to 135k governed experience cards.

\section{Related Work}

\textbf{Memory Construction for Agents.}
Building effective memory systems requires ensuring data quality and standardization. Early approaches mined fix patterns from commits or patches \cite{pan,kim2013automatic} but captured only syntactic transformations. 
Recent agent memory systems have explored structured experience organization. ExpeRepair \cite{mu2025experepair} maintains dual memory banks separating episodic demonstrations from semantic insights, while SWE-Exp \cite{Swe-exp} extracts multi-level experiences from agent trajectories.

\textbf{Code Agents.}
Recent advances in LLMs have driven autonomous code agents for automated software engineering. SWE-agent \cite{Swe-agent} pioneered Agent-Computer Interfaces with specialized tools for LLM-based code navigation, while AutoCodeRover \cite{autocoderover} enhanced fault localization through syntax tree representations and spectrum-based techniques. Agentless \cite{agentless} demonstrated that simpler pipeline approaches can match agent performance with lower costs. Training-based methods have also emerged: SWE-Fixer \cite{Swe-fixer} compiled 110K instances for fine-tuning open-source repair models, BugPilot \cite{bugpilot} generated synthetic bugs for efficient skill learning, and Co-PatcheR \cite{Co-PatcheR} explored modular patching with component-specific models. Existing approaches largely operate within single repositories or rely on self-generated experiences, overlooking the vast corpus of cross-repository human debugging expertise in GitHub's collaborative ecosystem.

In contrast, MemGovern addresses these limitations through comprehensive governance that curates cross-repository human experiences from GitHub at scale. Our approach introduces multi-stage curation to filter noisy discussions, content purification to standardize heterogeneous issue reports, and a dual-layer protocol that separates retrieval signals from actionable repair logic, enabling effective learning from the collective debugging knowledge embedded in real-world collaborative development.

\section{MemGovern}

\begin{figure*}[!t]
  \centering
  \includegraphics[width=\textwidth]{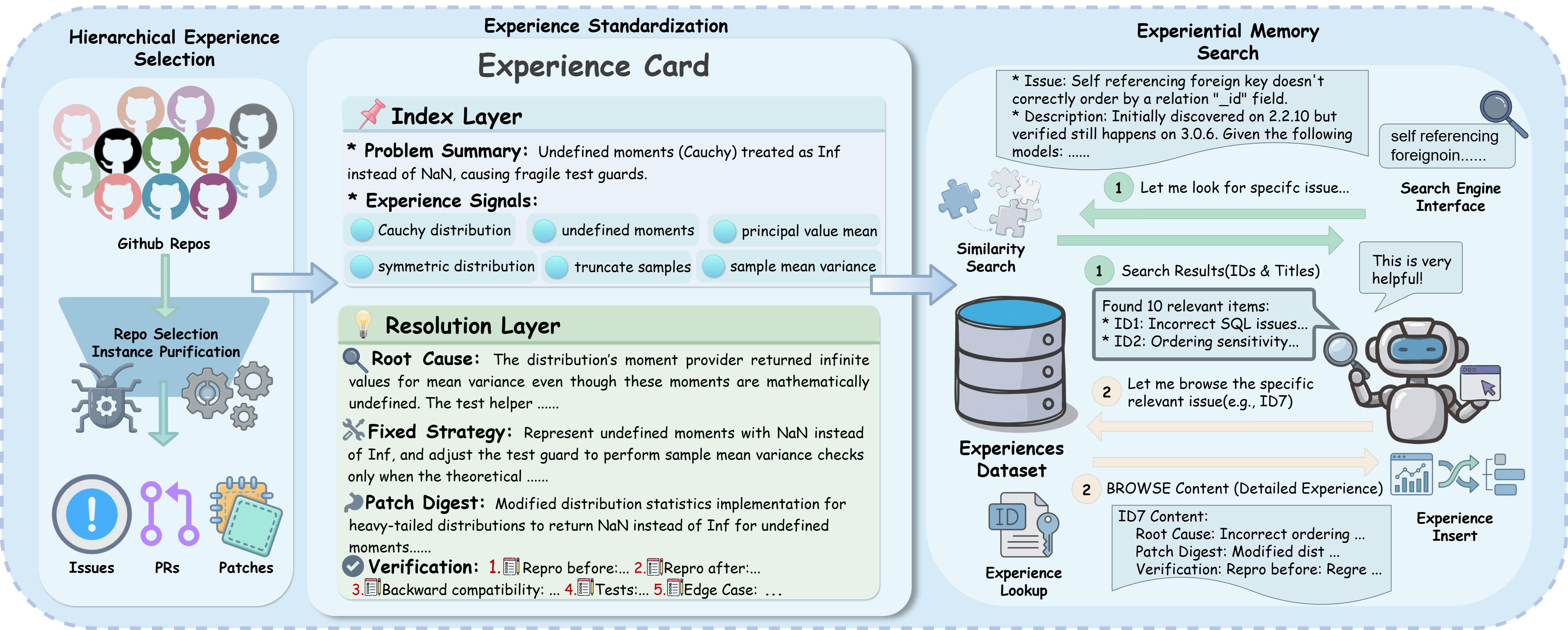}
  \caption{Architecture of MemGovern. MemGovern selects raw human experiences from GitHub and standardizes them into experience cards, enabling agents to utilize them through agentic experience search.}
  \label{fig:architecture}
  \vspace{-0.6em} 
\end{figure*}

In this paper, we propose MemGovern, a unified experience governance framework that transforms raw human experiences into a standardized, agent-friendly experiential memory , thereby allowing agents to learn from human experience in an effective and reliable manner. As illustrated in Figure~\ref{fig:architecture}, MemGovern constructs a large collection of experience cards through \textbf{\emph{experience governance}} and subsequently enables efficient utilization of human experience via \textbf{\emph{agentic experience search}}. The details are introduced as follows.

\subsection{Experience Governance} \label{subsec:governance}

The efficacy of experience-driven learning hinges on the quality of the underlying data. While GitHub offers a vast repository of debugging expertise, this open-world knowledge is inherently unstructured and noisy. Historical records are often laden with social chatter, ambiguous process descriptions, and non-standardized terminology, creating a "semantic gap" that hinders direct agentic retrieval. To bridge this gap, MemGovern involves \textbf{\textit{Experience Governance}}, a systematic pipeline that distills chaotic raw data into structured, high-fidelity \textbf{\textit{experience cards}}. This process prioritizes information density and experience reliability through three stages: (1) \textit{Experience Selection}, which filters low-signal noise at both the repository and instance levels; (2) \textit{Standardization}, which reconstructs raw data through a Unified repair-experience protocol; and (3) \textit{Quality Control}, a checklist-based mechanism to ensure memory fidelity.

\subsubsection{Hierarchical Experience Selection}
\label{sub:selection3.1.1}
The primary challenge in building a reliable experiential memory is the high quality variance, ranging from unmaintained hobby projects to knowledge-dense issues. Integrating such low-signal data risks memory pollution. To address this, MemGovern uses a two-tiered selection strategy that ensures both source authority and instance completeness.

\textbf{Repository Selection.} To ensure the agent learns from active and high-quality software engineering practices, MemGovern filters for repositories exhibiting sustained maintenance. MemGovern selects top-$M$ repositories based on a score that balances popularity (Stars $S_r$) with maintenance intensity (Issues $I_r$ and Pull Requests $P_r$):
\begin{gather}
\begin{aligned}
\mathrm{Score}(r)=\lambda_s \log(1+S_r)+\lambda_i \log(1+I_r)+\lambda_p \log(1+P_r),
\end{aligned}
\end{gather}
where $\lambda_s, \lambda_i, \lambda_p$ represent weights for stars, issues, and PRs respectively. This ensures the source data reflects robust, active development flows.

\textbf{Instance Purification.} Within selected repositories, MemGovern rigorously filters $(Issue, PR, Patch)$ triplets to retain only "closed-loop" repair records. A valid instance must provide a complete chain of evidence: explicit linkage between the issue and merged code, a parsable diff, and diagnostic anchors (e.g., stack traces). Furthermore, to strip away social noise, MemGovern analyzes the comment threads and discard instances where the technical-content ratio falls below a threshold $\tau=0.2$, preserving only threads rich in debugging logic.

\subsubsection{Experience Standardization}
\label{sub:standardization}
Raw GitHub discussion threads are inherently verbose, often interleaving failure symptoms with implementation details, social exchanges, and repository-specific context. Such entanglement obscures the essential signals required for effective retrieval. To render these records actionable, MemGovern defines a \textbf{\textit{unified repair experience protocol}} and standardizes selected instances accordingly.

The core philosophy of this protocol is to explicitly decouple retrieval semantics from reasoning logic. Concretely, MemGovern begins with \textbf{\textit{content purification}}, where an LLM compresses the original comment stream by removing non-technical interactions (e.g., greetings, merge notifications) and redundant execution logs. The purified content is then reorganized into two functionally distinct layers within each experience card: the \textit{Index Layer} and the \textit{Resolution Layer}.

\textbf{Index Layer.}  
This layer captures the information available at an agent’s initial observation stage and serves as the primary retrieval interface. It includes a normalized Problem Summary and a set of generalizable Diagnostic Signals (e.g., exception types, error signatures, component-level tags). Repository-specific identifiers and incidental implementation details are intentionally removed to maximize semantic matchability across heterogeneous repositories.

\textbf{Resolution Layer.}  
This layer encapsulates the transferable repair knowledge distilled from human debugging processes. It contains the Root Cause analysis, an abstract Fix Strategy, and a concise Patch Digest. By isolating causal reasoning and procedural patterns from concrete code artifacts, this layer enables agents to reuse human repair logic beyond the original context.
Formally, each standardized experience card $E_i$ is represented as:
\begin{gather}
E_i = \langle \text{Index}=I_i,\ \text{Resolution}=R_i \rangle.
\end{gather}
where $I_i$ denotes the index layer with retrievable failure symptoms, and $R_i$ denotes the resolution layer with reusable repair logic. This structural decoupling allows agents to retrieve experiences based on symptom-level similarity encoded in the Index Layer, while executing repairs using abstract reasoning strategies encoded in the Resolution Layer, thereby enabling effective cross-repository generalization.

\subsubsection{Checklist-Based Quality Control}
\label{sub:qc}

Automated extraction pipelines can occasionally hallucinate details or miss key nuances. To guarantee the reliability of the memory bank, we introduce a checklist-based quality control mechanism that acts as a final gatekeeper before ingestion.

We employ an LLM as a structured evaluator to score each generated experience card against critical dimensions. Unlike simple filtering, this process incorporates a \textit{Refine Loop}: if a card's aggregate score falls below a threshold $\gamma$, the evaluator provides targeted feedback on the deficient dimensions. The extraction pipeline then regenerates only the problematic sections based on this feedback. This cycle repeats for a maximum of three iterations, ensuring that only experiences passing this rigorous quality gate are indexed. Consequently, MemGovern operates on a foundation of verifiable expert knowledge rather than noisy raw data.

\subsection{Experiential Memory Search}
\label{sec:agent_method}
Constructing a high-quality experiential memory  is only half the challenge; the agent must also be able to retrieve and apply this experience effectively. Unlike standard Retrieval-Augmented Generation (RAG), which relies on single-shot context injection, real-world debugging is a dynamic process of hypothesis formulation and validation. To support this process, MemGovern adopts an agentic experience search mechanism that mirrors how human engineers navigate technical documentation. Specifically, MemGovern introduces a dual-primitive interface that enables agents to interact with a well-governed experiential memory  and complete complex coding tasks through progressively agentic search over stored experiences.

\subsubsection{Dual-Primitive Interface}
To balance the need for broad semantic discovery with the cost of detailed context processing, MemGovern exposes the experiential memory to agent through two primitives: \textit{Searching} and \textit{Browsing}.

\textbf{Searching.}
The Searching tool serves as a high-throughput filter that enables the agent to scan the memory index using a query. Given a task-specific query $q$, the tool traverses the index to identify relevant experience cards. Retrieval is based on the Index Layer of each card, ensuring that matches are driven by symptom-level similarity rather than implementation details. The relevance of an experience card $E_i$ is computed via cosine similarity in the embedding space:
\begin{equation}
\mathrm{sim}(q,I_i)=\frac{\phi(q)\cdot \phi(I_i)}{\|\phi(q)\|\,\|\phi(I_i)\|}
\end{equation}
where $\phi(\cdot)$ is embedding function and $I_i$ represents index layer of card $E_i$. The tool returns a ranked list of candidates $\{(E^{(k)}, \mathrm{sim}(q,I^{(k)}))\}_{k=1}^{K}$, allowing the agent to assess relevance before proceeding to a deeper inspection.

\textbf{Browsing.}
After promising candidates are identified, the Browsing tool grants access to the detailed resolution layer of a selected card. It offers high-precision retrieval without overwhelming the agent with raw details. This design ensures that once a relevant case is identified, the agent receives a clear blueprint of \textit{what} changed and \textit{why}, enabling direct transfer of human expertise to the agent.

\subsubsection{Progressive Agentic Search}

Building upon these primitives, we introduce a \emph{progressive agentic search} mechanism that adapts dynamically to the evolving problem-solving state. Unlike conventional approaches that follow a rigid, predefined pipeline, the agent continuously reassesses its informational needs and autonomously choosing searching and browsing actions.

\input{Tables/main-results}

\textbf{Query Formulation and Retrieval.}
The process begins with problem representation. The agent analyzes the current issue to extract diagnostic keywords, such as failure symptoms, failing test names, stack traces, and relevant module identifiers, to construct an initial query $Q$. It then invokes the \textit{Searchings} tool to retrieve semantically related candidates. Rather than indiscriminately consuming all results, the agent autonomously assesses the returned metadata and selects the most promising subset for deeper inspection. When the returned candidates are not sufficiently relevant, the agent can revise the query by incorporating additional observed symptoms and re-invoke \textit{Searching}, enabling multi-round retrieval without prematurely committing to noisy details.

\paragraph{Selective Browsing and Analogical Transfer.}
For the selected candidates, the agent employs the \textit{Browsing} tool to access their resolution layers. The core challenge here is \textit{analogical transfer}: the agent must map the historical solution to the current repository's context. By synthesizing the browsed evidence, the agent induces a transfer triplet: \textit{Root Cause Pattern} $\rightarrow$ \textit{Modification Logic} $\rightarrow$ \textit{Validation Strategy}. For instance, if a retrieved experience suggests "adding a boundary check for null inputs in the parser", the agent maps this abstract strategy to the specific variable names and API versions of the current codebase. This results in a concrete, actionable repair plan involving specific file locations and code edits. Totally, progressive agentic search enables the agent to autonomously and deeply acquire and exploit experiential memory.

\section{Experiments}

\subsection{Experimental Setup} 

We evaluate our method on SWE-bench Verified. For a comprehensive and fair evaluation, we compare MemGovern against SWE-Agent and several strong baselines. Due to space limitations, we provide a detailed introduction to the benchmark and baselines in Appendix \ref{app:exp}.

\subsection{Configuration}

During the bug-fix experience governance process, we collected active, well-maintained open-source repositories from GitHub with more than 100 stars. After verification and data cleaning, we obtained approximately 150K Issue–PR–Patch triplets. These triplets were then processed by our experience governance pipeline using GPT-5.1 (medium reasoning), as described in Section~\ref{subsec:governance}. After de-duplication, the final experience card collection contains 135K items.

Building on the SWE-Agent framework, we implemented two tools: an Experience Search tool and an Experience Browse tool. Both tools are integrated into SWE-Agent’s default toolset, together with the anthropic style code editor and patch submission module. The Experience Search tool takes a search query and a hyperparameter, top-k, which controls the number of experience-card previews returned (default by 10 to balance context length and information gain). The Experience Browse tool takes the unique index of an experience card and returns the full bug-fix experience.

\begin{figure*}[t!]
    \centering
    \begin{minipage}[c]{0.55\textwidth}
        \centering
        \includegraphics[width=0.95\linewidth]{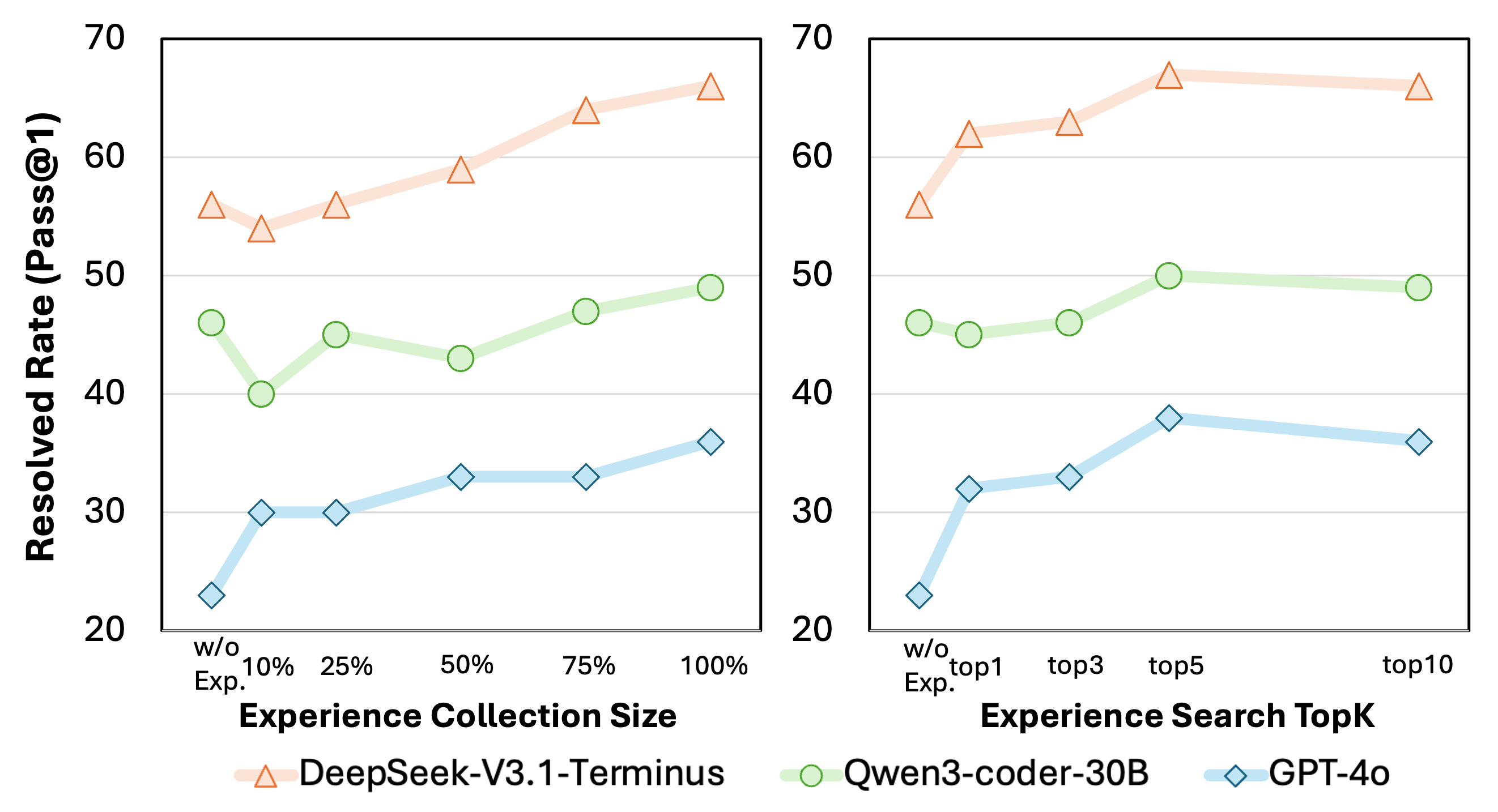}
        \caption{Results of MemGovern under various experiential memory size and MemGovern under various retrieval sizes.}
        \label{fig:experience_size_topk}
    \end{minipage}%
    \hfill 
    \begin{minipage}[c]{0.42\textwidth}
        \centering
        \small 
        \setlength{\tabcolsep}{2.5pt} 
        
        \begin{tabular}{llccc}
            \toprule
            \textbf{LLM} & \textbf{Strategy} & \textbf{Exp} & \makecell{\textbf{Resolved}\\ \textbf{Rate}} & $\Delta$ \\
            \midrule
            \multirow{4}{*}{\makecell[l]{DeepSeek-\\ V3.1-T}}
                & Base Agent    & $\times$     & 62.8 & -   \\
                & w/ RAG        & $\checkmark$ & 64.4 & +1.6 \\
                & w/ Ag. RAG    & $\checkmark$ & 63.4 & +0.6 \\
                & w/ Ag. Search & $\checkmark$ & \textbf{65.8} & +3.0 \\
            \midrule
            \multirow{4}{*}{\makecell[l]{Qwen3-\\ Coder}}
                & Base Agent    & $\times$     & 48.0 & -   \\
                & w/ RAG        & $\checkmark$ & 46.8 & -1.2 \\
                & w/ Ag. RAG    & $\checkmark$ & 48.6 & +0.6 \\
                & w/ Ag. Search & $\checkmark$ & \textbf{51.4} & +3.4 \\
            \midrule
            \multirow{4}{*}{GPT-4o}
                & Base Agent    & $\times$     & 23.2 & -   \\
                & w/ RAG        & $\checkmark$ & 31.2 & +8.0 \\
                & w/ Ag. RAG    & $\checkmark$ & 31.2 & +8.0 \\
                & w/ Ag. Search & $\checkmark$ & \textbf{32.6} & +9.4 \\
            \bottomrule
        \end{tabular}
        
        \captionof{table}{Comparison of experience usage strategies. ``Exp'' denotes experiential memory. ``Ag.'' denotes ``Agentic''.}
        \label{tab:ablation_retrieval}
    \end{minipage}
\end{figure*}

\subsection{Main Results}

Table~\ref{tab:main} presents the main results on SWE-bench Verified, comparing MemGovern with the SWE-Agent baseline across seven LLMs. MemGovern consistently outperforms the baseline across all models, demonstrating robust and model-agnostic improvements. The gains are especially pronounced for models with weaker initial performance, while even the stronger models benefit noticeably, highlighting MemGovern’s ability to enhance task-solving capabilities across a diverse range of LLMs. Although MemGovern introduces some additional token overhead due to its memory mechanism, this cost is outweighed by the average performance improvement of around 4.65\%, making it a highly acceptable trade-off.

\section{Analysis}

\subsection{Ablation of Experience Governance}

To study the effectiveness of experience governance in MemGovern, we conduct ablation studies on two fundamental characteristics of governed experience: \emph{size} and \emph{quality}.

\textbf{Effect of Experiential Memory  Size.}
MemGovern adopts an experience-driven agentic framework with a governed \emph{experiential memory}. To verify whether MemGovern’s performance gains stem from broader coverage of governed human experience, we evaluate the MemGovern under various experiential memory  size. Specifically, as shown in Figure~\ref{fig:experience_size_topk}, we evaluate MemGovern under 10\%-100\% of the full experiential memory  (135K experience cards).
The results show that enlarging the experiential memory  steadily improves task resolution performance across LLMs. This improvement can be attributed to a higher probability of retrieving relevant and transferable experience cards when the memory coverage increases. Importantly, the gains are monotonic rather than abrupt, indicating that MemGovern does not rely on a small subset of exceptional examples but instead benefits cumulatively from diverse, governed experiences. Overall, MemGovern effectively leverages large-scale experience aggregation, validating the necessity of constructing a sizable experiential memory  through systematic governance.

\input{Figures/patch_or_experience_cases}

\textbf{Effect of Experiential Memory  Quality.}
MemGovern further incorporates \emph{Experience Standardization} and \emph{Quality Control} to ensure that the experiential memory  is not only large but also reliable and agent-friendly. To assess whether the observed gains truly arise from governed experience representations rather than raw historical data, we conduct an analysis experiment that ablates the quality of the experiential memory .
Concretely, as reported in Figure~\ref{fig:patch_or_experience_cases} we compare MemGovern using the same experiential memory  size but different memory contents: (i) raw \textit{PR+Patch} records without standardization or quality filtering, and (ii) fully governed experience cards constructed by MemGovern through the governance pipeline (see Section~\ref{subsec:governance}).
The results indicate that while raw records can occasionally provide partial benefits, their effectiveness is unstable and model-dependent, likely due to noise, verbosity, and weak alignment with agent retrieval semantics. In contrast, governed experience cards consistently yield stronger and more stable improvements across models. These findings show that MemGovern’s improvements are driven by experience governance rather than sheer data exposure, and they empirically validate the effectiveness of standardization and quality control in constructing a high-fidelity experiential memory .

\subsection{Superiority of Agentic Search}

MemGovern adopts an agentic experience search mechanism based on a dual-primitive interface. To verify that MemGovern’s performance gains stem from the proposed agentic search paradigm rather than from naive experience injection, we compare different experience utilization strategies. Specifically, as shown in Table~\ref{tab:ablation_retrieval}, we evaluate three variants under the same experiential memory : (i) \textbf{RAG}, which performs a single retrieval before the repair process and directly injects all retrieved experiences into the context; (ii) \textbf{Agentic RAG}, which allows retrieval to be triggered adaptively during iterative debugging but still follows a retrieve-and-inject paradigm; and (iii) \textbf{Agentic Search}, which decouples candidate discovery from evidence usage by first retrieving a broader candidate set and then selectively browsing and transferring only relevant experience cards. 

The results show that Agentic Search consistently outperforms both static and adaptive RAG variants across diverse backbone models, including DeepSeek-V3.1-Terminus, Qwen3-Coder-30B, and GPT-4o. This improvement is not merely a result of accessing a larger pool of experiences, but stems from the agent’s strategic management of searching depth and breadth. By decoupling \textbf{searching breadth} (broad candidate discovery) from \textbf{browsing depth} (selective evidence grounding and logic extraction), the agent can explore a wide range of candidate experiences without being overwhelmed by irrelevant information, while simultaneously diving deeply into the most promising ones leads to extract actionable insights. This allows reasoning over multiple experiences in a structured and iterative manner, ensuring that only contextually relevant knowledge is integrated with the current debugging state. In contrast, both RAG variants tend to conflate breadth with depth, making them highly sensitive to noise: weakly related experiences are injected directly into the context, often disrupting the generation process. Overall, our proposed agentic experience search is highly effective, enabling principled exploration of broad and deep experience spaces essential for complex software debugging.

\subsection{Effect of Various Retrieval Sizes}

Furthermore, to study how the number of retrieved candidates affects performance, we conduct a ablation study on the retrieval Top-$K$ parameter in the agentic search pipeline. Specifically, as reported in Figure~\ref{fig:experience_size_topk}, we show The performance of MemGovern under various Top-$K$ retrieved candidates.

The results reveal a clear and consistent trend across different backbone models. When Top-$K$ is small, increasing the retrieval size leads to steady performance improvements, indicating that a larger candidate pool increases the likelihood of surfacing actionable and relevant experiences. However, beyond a moderate value of $K$, the gains gradually diminish and eventually plateau.This behavior aligns well with the design of agentic experience search. At small $K$, retrieval coverage is the primary bottleneck, and expanding the candidate set helps the agent discover useful analogies. Once sufficient coverage is achieved, further enlarging the candidate pool yields limited benefits, as the agent already has access to representative experience patterns and additional candidates tend to be redundant. Importantly, performance does not degrade at larger $K$, which suggests that the selective browsing mechanism effectively filters out irrelevant experiences and prevents context overload. Therefore, MemGovern is robust to the choice of retrieval size and that a moderate Top-$K$ achieves a favorable balance between effectiveness and efficiency. These findings further support the effectiveness of the decoupled retrieval-and-browsing design, showing that agentic search can safely leverage broader retrieval without sacrificing reliability.

\begin{figure*}[t!]
    \centering
    \begin{minipage}[c]{0.60\textwidth}
        \centering
        \includegraphics[width=\linewidth]{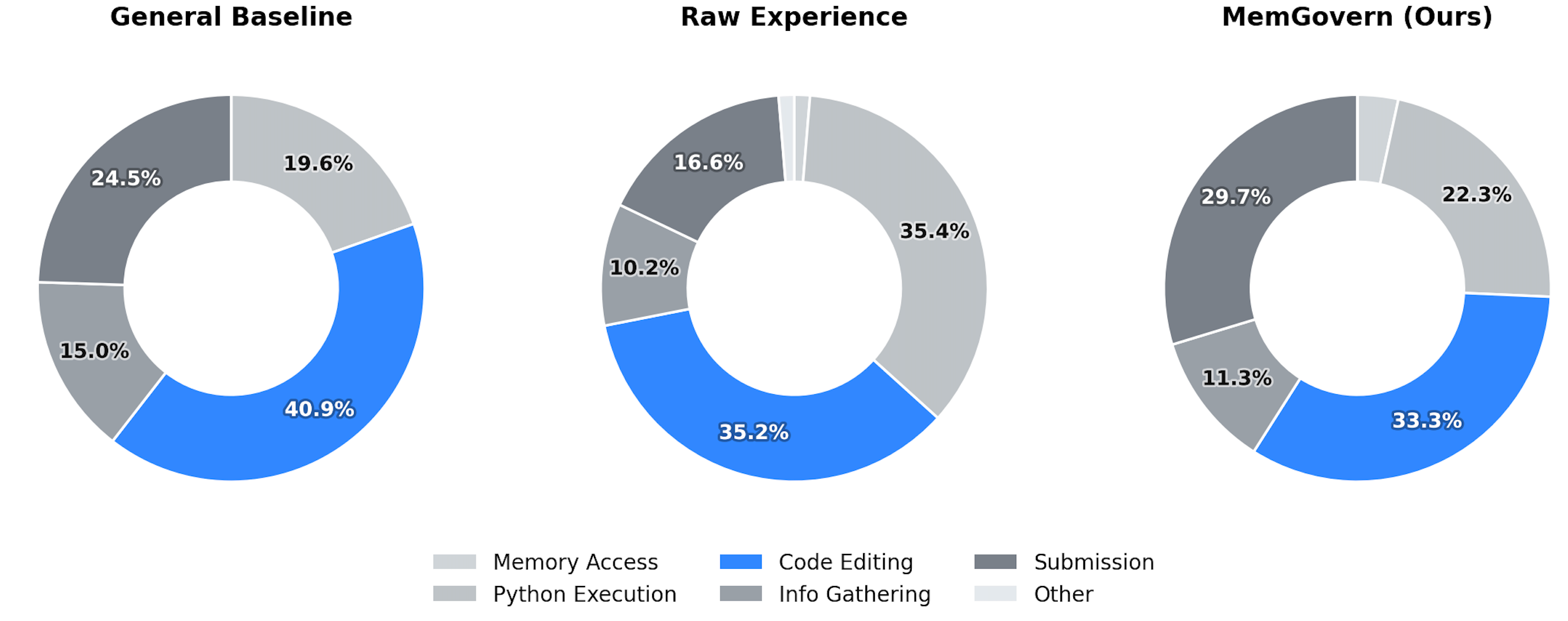}
    \end{minipage}%
    \hfill
    \begin{minipage}[c]{0.36\textwidth}
        \centering
        \raisebox{10pt}{\includegraphics[width=\linewidth]{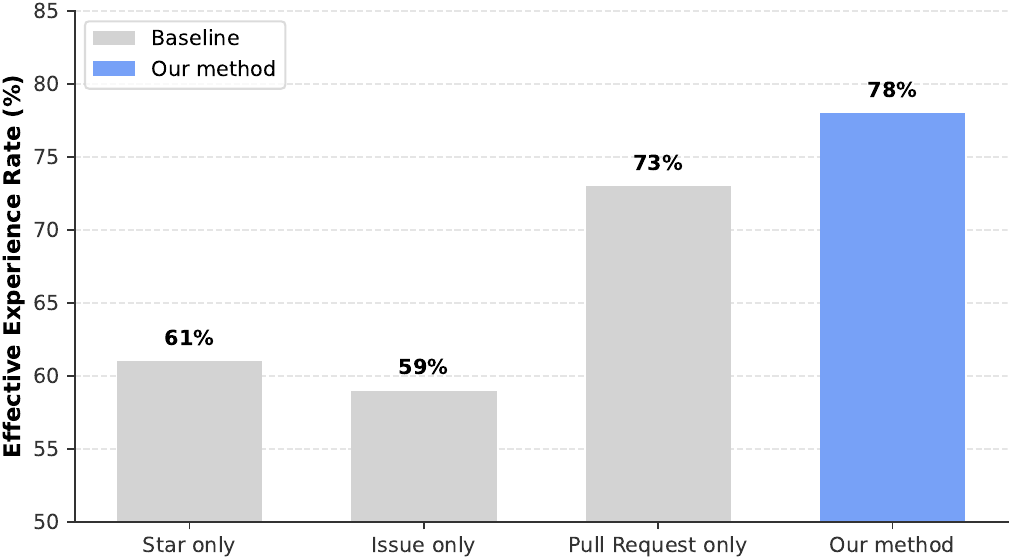}}
    \end{minipage}

    \vspace{2pt} 

    \begin{minipage}[t]{0.60\textwidth}
        \caption{Action composition of the agent across different experience settings. We report the fraction of steps spent on code editing versus other actions.}
        \label{fig:agent_action}
    \end{minipage}%
    \hfill
    \begin{minipage}[t]{0.36\textwidth}
        \caption{Human evaluation of repository selection. Proportion of issues judged reusable as transferable experience across different ranking rules.}
        \label{fig:effective_exp_rate}
    \end{minipage}
\end{figure*}

\subsection{Effect of Repository Selection}
To validate our repository selection strategy, we conduct a human evaluation on 100 randomly sampled GitHub repositories with $Sr \ge 100$ and similar Issue volumes. We then construct four Top-10 subsets ranked by \textit{Star only}, \textit{Issue only}, \textit{Pull Request only}, and \textit{Our method} (Section~\ref{sub:selection3.1.1}). Two experienced software engineers then assess whether issues in each subset can be distilled into transferable, reusable experiences. As shown in Figure~\ref{fig:effective_exp_rate}, our method yields the highest proportion of issues deemed reusable as experience, suggesting that combining popularity and maintenance signals selects repositories with denser, higher-quality experience candidates and strengthens the memory.

\subsection{Can Agent Learn from Experience?}

We further analyze agent behaviors by quantifying the proportion of each operation category (Figure~\ref{fig:agent_action}). The results indicate that MemGovern improves both success rate and efficiency by reducing unguided exploration and encourages effective self-testing.

Specifically, Info Gathering drops from 15.0\% to 11.3\% under MemGovern, indicating that experience cards offer targeted navigational guidance and reduce extensive repository exploration. This improvement aligns with a more balanced editing–execution strategy: the baseline agent often edits prematurely before adequately reproducing or localizing faults (Code Editing: 40.9\%; Code Execution: 19.6\%), raising the risk of faulty fixes. With MemGovern, the agent shifts focus toward self-testing (Code Execution: 22.3\%) and curtails unnecessary edits (Code Editing: 33.3\%) by leveraging root-cause and verification cues from experience cards, resulting in more reliable repairs.

Besides, although Raw Experience (unprocessed GitHub PR and patches) provides useful cues, it incurs a substantially higher verification overhead (35.4\% Code Execution), revealing that unrefined experience introduce noise and lead to low-quality fix. MemGovern mitigates this issue through knowledge governance, maintaining effective information while compressing irrelevant contexts.

\begin{figure*}[t]
    \centering
    \includegraphics[width=\linewidth]{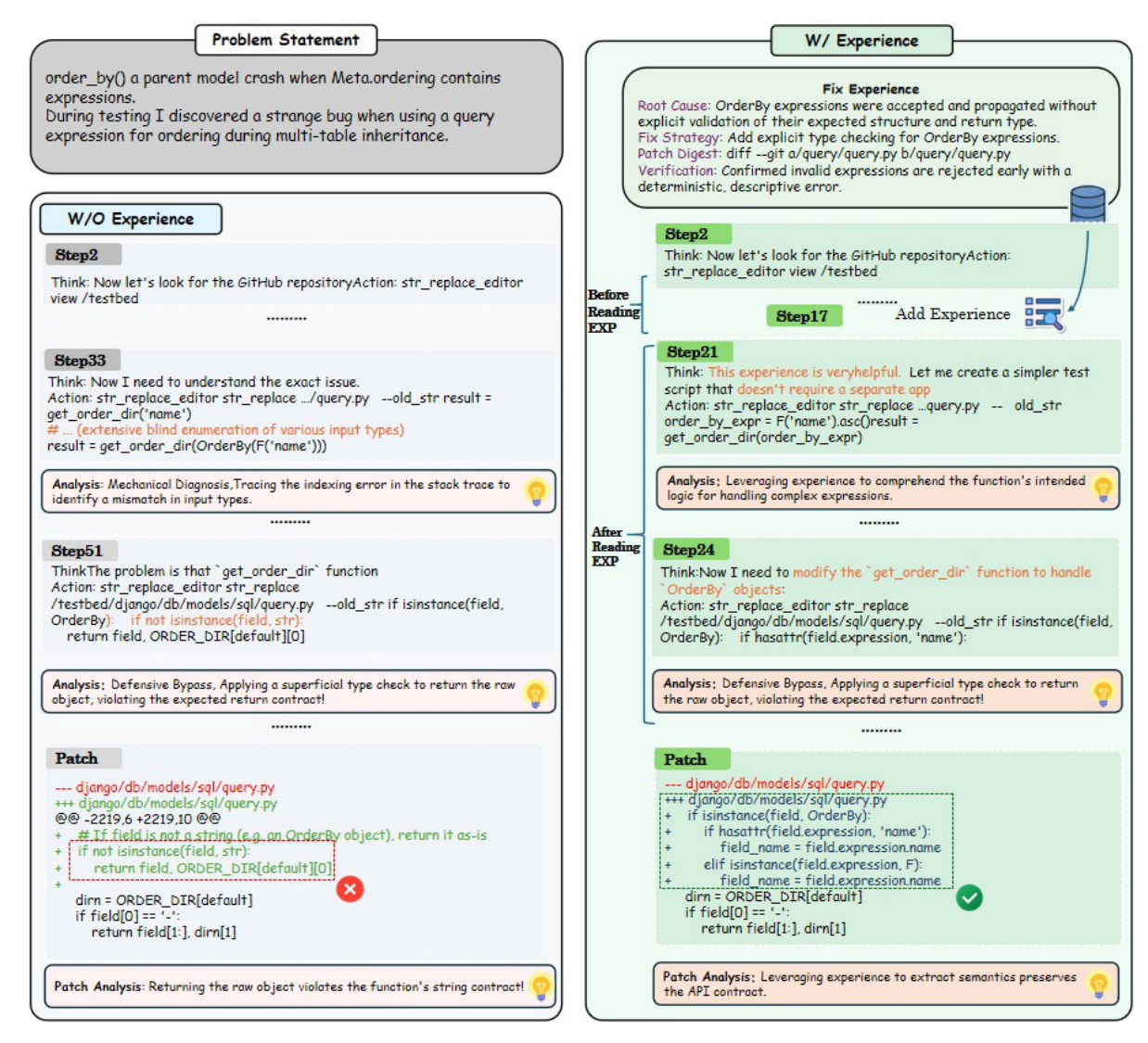}
    \caption{Comparison of Agent behavior without experience (Top) vs. with MemGovern's experiential memory (Bottom). While the baseline applies a defensive bypass that violates the API contract, MemGovern leverages historical logic to implement a semantically correct fix.}
    \label{fig:case_baseline}
\end{figure*}

\subsection{Case Study}
\label{sec:case_study}

To provide a qualitative understanding of how MemGovern enhances autonomous software engineering, we examine how the presence of experience enables the agent to move beyond superficial fixes to robust solutions.


Figure~\ref{fig:case_baseline} illustrates a case involving a crash in Django's \texttt{order\_by()} function when \texttt{Meta.ordering} contains query expressions.

\textbf{Without Experience (Baseline).} The baseline agent, lacking historical context, relies on a ``Mechanical Diagnosis.'' It traces the indexing error in the stack trace and attempts to fix the crash by blindly enumerating input types. This results in a ``Defensive Bypass'' patch: it applies a superficial type check (\texttt{if not isinstance(field, str)}) to return the raw object. While this suppresses the immediate crash, it violates the expected return contract of the function, potentially causing silent failures downstream.

\textbf{With Experience (MemGovern).} MemGovern retrieves a relevant experience card and leverages its structured \textbf{Resolution Layer} to guide the repair. Specifically, the \textbf{Root Cause} field informs the agent that ``OrderBy expressions were accepted... without explicit validation'', while the \textbf{Fix Strategy} provides a clear directive to ``add explicit type checking'' rather than bypassing the object. Leveraging these governed insights, MemGovern implements a semantic-aware fix that extracts the field name (\texttt{field.expression.name}), thereby preserving the API contract. This demonstrates how MemGovern's structured schema—decoupling diagnostic signals from actionable resolution logic—provides the precise reasoning necessary to distinguish between a superficial symptom-fix and a correct repair.

We provide further analysis on the superiority of agentic search and the necessity of experience governance in Appendix~\ref{app:case_study}.

\section{Conclusion}

In this paper, we propose MemGovern, a framework for governing raw GitHub data into agent-friendly experiential memory, along with a corresponding agentic experience search method that provides memory infrastructure for agents. MemGovern constructs 135k experience cards, and improves 4.65\% resolution rates on the SWE-bench, demonstrating its superior performance.

\section*{Limitations}
MemGovern is an agent-friendly memory infrastructure that enables agents to better learn from human experiences, thereby achieving improved performance. A limitation of MemGovern lies in the additional tokens generated when searching the memory during agent execution. Inevitably, the agent processes more tokens during searching in exchange for higher resolution rates. Overall, considering that MemGovern achieves an average 4.65\% improvement in resolution rates compared to SWE-Agent, we believe the slightly higher token consumption is acceptable. Strategies for compressing memory length will be explored in future work.

\newpage

\bibliography{custom}

\newpage

\appendix
\input{Appendix/A_Experimental_Setup}

\input{Appendix/B_Case_Study_Analysis}

\input{Appendix/C_Experience_Card_Schema}

\end{document}

%% file: Tables/main-results.tex
\begin{table*}[t]
\centering\small
\begin{tabular}{llcccc}
\toprule
\textbf{Method} & \textbf{LLM} & \textbf{Resolved Rate (\%)} & \textbf{$\Delta$ (\%)} & \textbf{Avg. Tokens (M)} & \textbf{Avg. Cost (\$)} \\
\midrule
\multicolumn{6}{l}{\textbf{Other baselines}} \\
\midrule
AutoCodeRover   & GPT-4o                    & 28.8 & -    & - & - \\
CodeAct         & GPT-4o                    & 30.0 & -    & - & - \\
SWESynInfer     & GPT-4o                    & 31.8 & -    & - & - \\
mini-SWE-agent  & DeepSeek-V3.2-Reasoner    & 60.0 & -    & - & - \\
mini-SWE-agent  & Claude-4-Opus             & 67.6 & -    & - & - \\
mini-SWE-agent  & GPT-5.2                   & 69.0 & -    & - & - \\
\midrule
\multicolumn{6}{l}{\textbf{SWE-Agent vs. MemGovern across LLMs}} \\
\midrule
SWE-Agent  & \multirow{2}{*}{\begin{tabular}[c]{@{}l@{}}Claude-4-Sonnet\end{tabular}} & 66.6 & -    & 2.29 & 6.94 \\
MemGovern  &  & \textbf{69.8} & +3.2 & 2.35 & 7.27 \\
\midrule
SWE-Agent  & \multirow{2}{*}{\begin{tabular}[c]{@{}l@{}}GPT5-Medium\end{tabular}} & 65.0 & -    & 0.78 & 1.14 \\
MemGovern  &  & \textbf{67.4} & +2.4 & 0.91 & 1.33 \\
\midrule
SWE-Agent  & \multirow{2}{*}{\begin{tabular}[c]{@{}l@{}}DeepSeek-V3.1T\end{tabular}} & 62.8 & -    & 0.89 & 0.18 \\
MemGovern  &  & \textbf{65.8} & +3.0 & 0.95 & 0.19 \\
\midrule
SWE-Agent  & \multirow{2}{*}{\begin{tabular}[c]{@{}l@{}}Qwen3-235B\end{tabular}} & 47.2 & -    & 1.17 & 0.09 \\
MemGovern  &  & \textbf{55.4} & +8.2 & 1.42 & 0.10 \\
\midrule
SWE-Agent  & \multirow{2}{*}{\begin{tabular}[c]{@{}l@{}}Kimi-K2-Instruct\end{tabular}} & 43.8 & -    & 0.57 & 0.26 \\
MemGovern  &  & \textbf{51.8} & +8.0 & 1.14 & 0.53 \\
\midrule
SWE-Agent  & \multirow{2}{*}{\begin{tabular}[c]{@{}l@{}}Qwen3-Coder-30B\end{tabular}} & 48.0 & -    & 0.92 & 0.07 \\
MemGovern  &  & \textbf{51.4} & +3.4 & 0.99 & 0.07 \\
\midrule
SWE-Agent  & \multirow{2}{*}{\begin{tabular}[c]{@{}l@{}}GPT-4o\end{tabular}} & 23.2 & -    & 0.73 & 1.38 \\
MemGovern  &  & \textbf{32.6} & +9.4 & 0.97 & 1.84 \\
\midrule
SWE-Agent  & \multirow{2}{*}{\begin{tabular}[c]{@{}l@{}}GPT-4o-Mini\end{tabular}} & 14.0 & -    & 0.99 & 0.14 \\
MemGovern  &  & \textbf{17.2} & +3.2 & 1.37 & 0.20 \\
\bottomrule
\end{tabular}
\caption{Comparisons with prior agents. $\Delta$ is computed against SWE-Agent under the same backbone.}
\label{tab:main}
\end{table*}

%% file: Figures/patch_or_experience_cases.tex
\begin{figure*}[t]
    \centering
    \includegraphics[width=\linewidth]{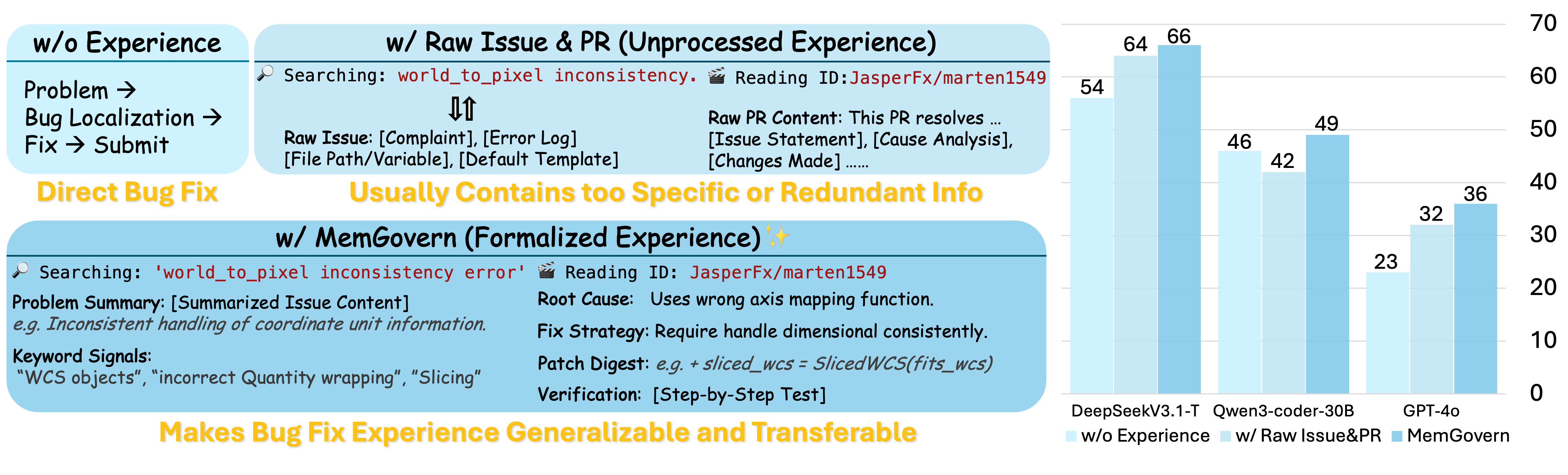}
    \caption{Comparison of using raw unprocessed experience and MemGovern's experience.}
    \vspace{-\textfloatsep}
    \vspace{10pt}
    \label{fig:patch_or_experience_cases}
\end{figure*}

%% file: Appendix/A_Experimental_Setup.tex
\section{Experimental Setup}
\label{app:exp}

\textbf{Benchmark.} We evaluate our method on SWE-bench Verified, a high-quality subset of SWE-bench containing 500 real-world GitHub issues, focused on functional bug fixing in a controlled, self-contained setup. For each instance, the model is given only the natural-language issue description and the corresponding repository. Correctness is assessed by running developer-written unit tests against the generated patch, providing a consistent and rigorous evaluation of automated bug-fixing performance.


\textbf{Baselines.} For a comprehensive and fair evaluation, we compare MemGovern against SWE-Agent and several strong baselines including SWE-Agent \cite{sweagent2024},  AppMap Navie \cite{appmapnavie}, AutoCodeRover \cite{autocoderover}, CodeAct \cite{codeact} and SWESynInfer \cite{SWESynInfer}. The comparison is conducted across multiple LLM backbones, covering both open and proprietary paradigms. Specifically, we evaluate four leading open-source models(DeepSeek-V3.1-Terminus, Qwen3-Coder-30B, Kimi-K2-Instruct, and Qwen3-235B) as well as five closed-source models(GPT-4o, GPT-4o mini, Claude Sonnet 4, GPT-5, and Gemini 3 Pro). Notably, MemGovern is implemented as a plug-and-play module that can be integrated into existing agent scaffolds with minimal changes; unless otherwise stated, we adopt SWE-Agent as the shared backbone for subsequent experiments.

%% file: Appendix/B_Case_Study_Analysis.tex
\section{Additional Case Study Analysis}
\label{app:case_study}

Following the case study in Section~\ref{sec:case_study}, we present two additional comparisons to highlight different aspects of MemGovern's performance: the superiority of agentic search over static retrieval, and the necessity of experience governance.

\subsection{Superiority of Agentic Experience Search}
\begin{figure*}[t]
    \centering
    \includegraphics[width=\linewidth]{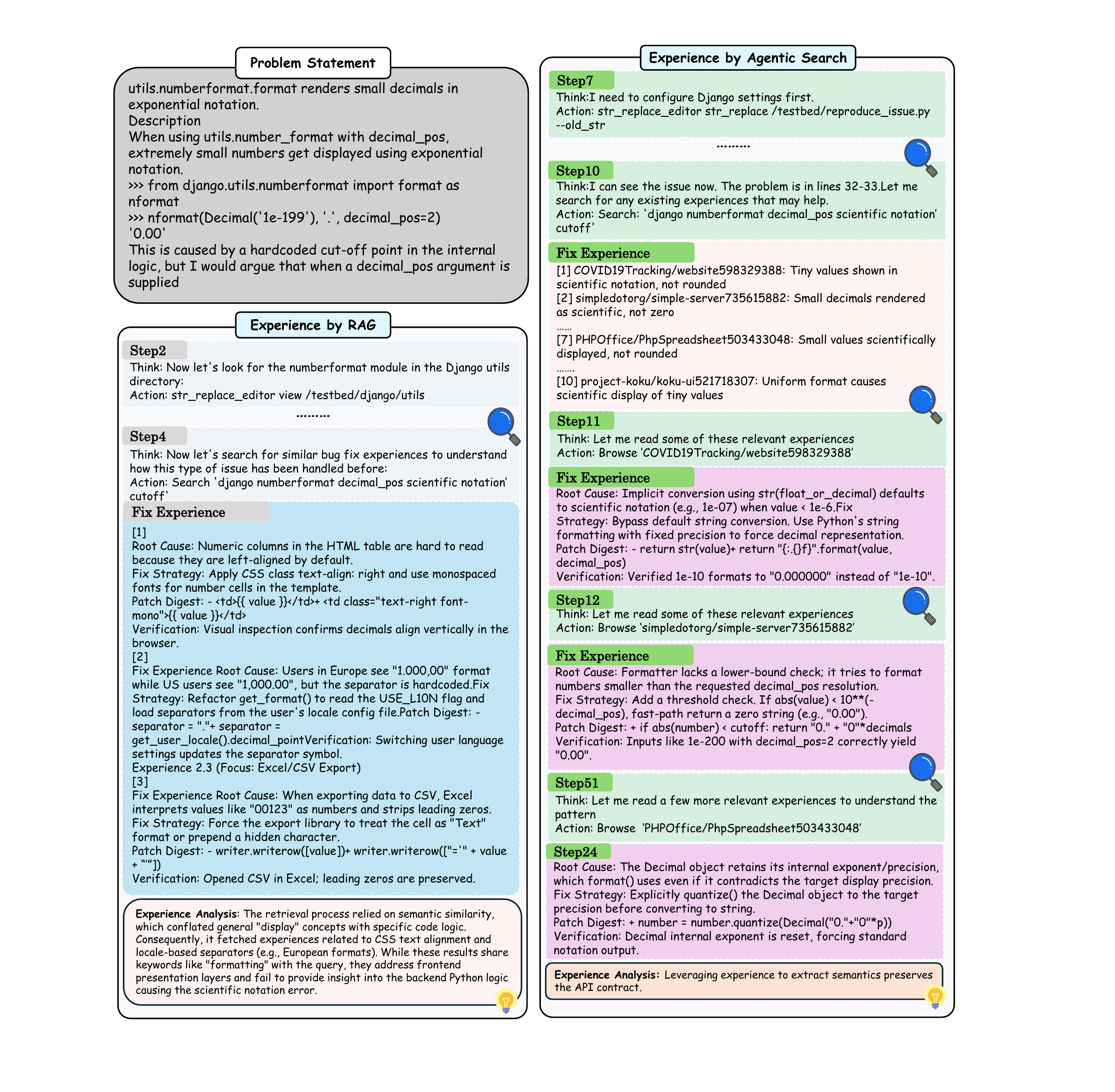}
    \caption{Comparison of Standard RAG (Top) vs. MemGovern's Agentic Search (Bottom). RAG retrieves irrelevant frontend formatting experiences due to surface similarity. MemGovern's agentic workflow allows it to browse and locate the specific backend logic for handling scientific notation thresholds.}
    \label{fig:case_rag}
\end{figure*}

Standard RAG often suffers from semantic drift, where retrieved documents share surface keywords but differ in technical substance. Figure~\ref{fig:case_rag} compares MemGovern's Agentic Search against a Standard RAG baseline on a Django issue where \texttt{number\_format} renders small decimals in scientific notation.

\textbf{Standard RAG.} As defined in our comparisons, the RAG baseline performs a single retrieval before the repair process and directly injects all retrieved experiences into the context. In this case, the retrieval is driven by surface similarity with the query ``django numberformat decimal\_pos,'' fetching irrelevant experiences related to frontend CSS alignment (e.g., ``text-align: right'') or locale-based separators. Influenced by this noise, the agent attempts a fix that aggressively mutates the number sign (\texttt{abs(number)}) early in the function, introducing a ``Defensive Bypass'' that breaks sign handling for other numeric types.

\textbf{Agentic Search (MemGovern).} In contrast, MemGovern employs a \textit{Search-then-Browse} workflow. The agent first retrieves candidates and then uses the \textit{Browsing} tool to inspect the logic. It successfully identifies a relevant case regarding ``scientific notation cutoff'' and rejects irrelevant frontend issues. By transferring the specific logic found in the experience—adding a threshold check, MemGovern produces a precise fix that resolves the issue without side effects. This highlights that the ability to iteratively browse and verify experience is crucial for filtering retrieval noise.

\subsection{Necessity of Experience Governance}
\begin{figure*}[t]
    \centering
    \includegraphics[width=\linewidth]{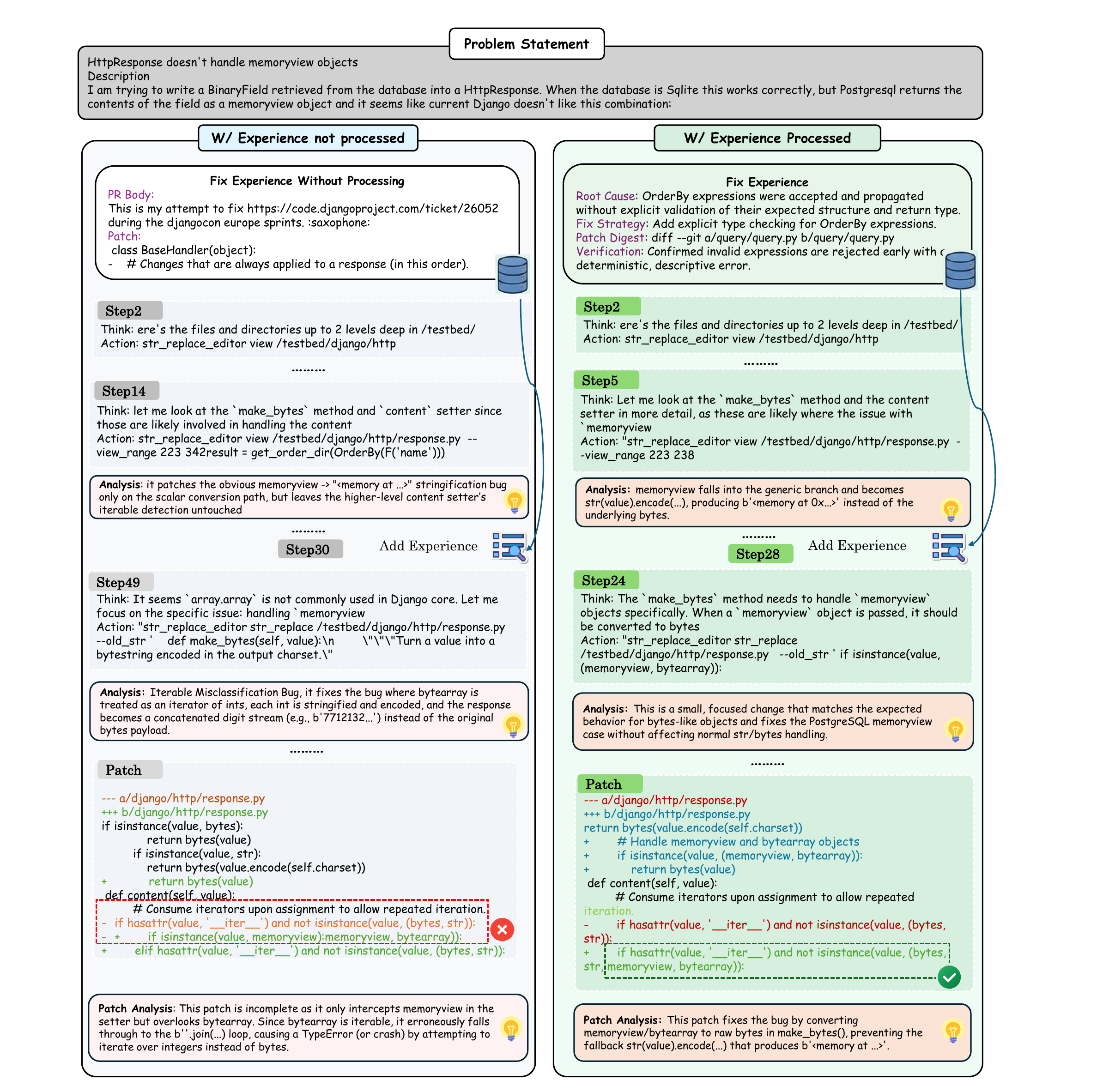}
    \caption{Comparison of Raw Experience (Left) vs. MemGovern's Processed Experience (Right). Raw experience distracts the agent with unrelated code changes, leading to an incomplete patch. Governed experience isolates the core repair logic, enabling the agent to correctly handle both \texttt{memoryview} and \texttt{bytearray} cases.}
    \label{fig:case_raw}
\end{figure*}

Finally, we analyze the importance of governing raw data into structured memory. Figure~\ref{fig:case_raw} depicts a complex issue involving \texttt{HttpResponse} handling \texttt{memoryview} objects from PostgreSQL.

\textbf{Raw PR+Patch Records.} When provided with raw PR+Patch records without standardization or quality filtering (Left), the agent is overwhelmed by noise. The retrieved raw data contains extraneous information about \texttt{BaseHandler} and \texttt{make\_bytes} that is not central to the specific \texttt{memoryview} edge case. Confused by the verbose diffs and social commentary, the agent misclassifies the issue as a generic iterable problem. It produces a patch that intercepts \texttt{memoryview} but fails to handle \texttt{bytearray} correctly, causing the code to fall through to a generic loop that crashes when iterating over bytes.

\textbf{Governed Experience Cards (MemGovern).} MemGovern (Right) utilizes fully governed experience cards where noise has been stripped via \textit{Content Purification}. The \textit{Resolution Layer} clearly highlights the specific fix strategy: ``Handle memoryview and bytearray objects.'' Freed from the distraction of unrelated code changes in the original PR, the agent focuses on the core logic. It implements a focused patch that explicitly converts both \texttt{memoryview} and \texttt{bytearray} to raw bytes, ensuring correct serialization. This confirms that experience governance is essential to convert noisy open-source data into actionable, agent-friendly memory.

\label{app:case}

%% file: Appendix/C_Experience_Card_Schema.tex

\definecolor{SchemaBg}{gray}{0.96}
\definecolor{SchemaBorder}{gray}{0.65}

\section{Experience Card Schema}
\label{app:experience-card-schema}

A detailed explanation of MemGovern's Experience Card Schema is described in Fig~\ref{fig:experience-card-schema}.

\begin{figure*}[!t]
\centering
\setlength{\fboxsep}{6pt}
\setlength{\fboxrule}{0.6pt}
\fcolorbox{SchemaBorder}{SchemaBg}{%
\begin{minipage}{\dimexpr\textwidth-2\fboxsep-2\fboxrule\relax}
\small

\medskip
\textbf{Fields}
\begin{itemize}
  \item \texttt{"Problem Summary"} \textit{(String)}: Concise technical summary of the bug root pattern. Must be generalizable; do NOT include repository names, commit hashes, or overly specific variable names.
  \item \texttt{"Signals"} \textit{(List[String], 10--18 items)}: High-signal matchable keywords/phrases extracted ONLY from the issue content. Each item is a short term (typically 2--4 words) covering mechanism/type, symptom/failure mode, trigger/input/environment, and affected component; include common aliases when helpful (e.g., \texttt{NPE} / \texttt{null pointer}). Avoid vague stop words; no duplicates.
  \item \texttt{"Root Cause"} \textit{(String)}: Evidence-backed causal chain from trigger to failure and the violated assumption. If uncertain, provide 1--2 candidate root causes, each with supporting evidence and a concrete observation/test to disambiguate.
  \item \texttt{"Fix Strategy"} \textit{(String)}: Design-level summary of how the bug was addressed (e.g., validation/guards, ordering/state transitions, parsing rules, synchronization, error handling/reporting). Mention constraints and compatibility notes (behavior/API, backward compatibility, performance impact), plus trade-offs/risks and mitigations. Do NOT restate the patch line-by-line.
  \item \texttt{"Patch Digest"} \textit{(String)}: Structured semantic digest grounded in the diff. Must include \textit{Changed Areas} (generalized components/modules/files) and 3--8 \textit{Key Chunks} describing what changed and why, linked back to the root cause. Do NOT paste raw diffs or large code blocks.
  \item \texttt{"Verification"} \textit{(String)}: Concrete, checkable verification plan and evidence: reproduce steps (before/after), tests added/updated (unit/integration/regression), and key edge cases/boundaries. If the patch has no tests, propose minimal steps with expected outcomes.
\end{itemize}
\end{minipage}%
}
\caption{MemGovern's Experience Card Schema.}
\label{fig:experience-card-schema}
\end{figure*}